# Optimal Allocation of Virtual Inertia Devices for Enhancing Frequency Stability in Low-Inertia Power Systems

Mingjian Tuo
*Student Member, IEEE*
Department of Electrical and Computer Engineering
University of Houston
Houston, TX, USA
mtuo@uh.edu

Xingpeng Li
*Member, IEEE*
Department of Electrical and Computer Engineering
University of Houston
Houston, TX, USA
xli82@uh.edu

***Abstract*—As renewable resources gradually replace conventional generation based synchronous machines, the dynamics of the modern grid changes significantly and the system synchronous inertia decreases substantially. This transformation poses severe challenges for power system stability; for instance, it may lead to larger initial rate of change of frequency and increase frequency excursions. However, new opportunities also arise as novel converter control techniques, so-called grid-forming strategies, show higher efficiency and faster response than conventional synchronous generators. They mainly involve virtual inertia (VI) emulation to mimic the behavior of synchronous machines. In this study, a state-space model for the power system network is developed with VI as a frequency regulation method. A reduced model based $H_2$-norm algorithm (RMHA) considering the Fiedler mode impact is proposed in this paper to optimize the allocation of VI devices and improve power system frequency stability. Finally, case studies conducted on the IEEE 24-bus system demonstrate the efficacy of the proposed RMHA approach.**

## I. Introduction

Modern power systems are required to accommodate an increasing volume of renewable energy sources. Increasing penetration of converter based renewable energy sources (RES) has introduced dynamic changes in modern power systems. Traditionally, the system inertia is primarily provided by the conventional synchronous generators. Due to the strong coupling between the synchronous generator's rotor and the power system, the inertia stored in synchronous generator rotor plays an important role in regulating the power system frequency dynamics. The system inertia can largely affect the initial rate of change of frequency (RoCoF) which manifests the dynamics between power and frequency during a short period of time following a power mismatch event.

Despite the coronavirus pandemic [1], the annual global renewable capacity addition increased by 45% to around 280 GW in 2020 [2]. In South Australia power system, the instantaneous penetration level of wind and solar capacity reaches 50% [3]. For Nordic, renewable generation has taken the place of nuclear power plants. A consequence of this transition is low inertia which is considered as one of the three main challenges faced by the system operator [4]. In the ERCOT system, wind power has rapidly developed over the last 20 years, and wind power accounted for 15% of the total generation in 2017 [5].

Different from synchronous generators, RESs are completely decoupled from the grid by the converter, making low to zero contributions to power system inertia. High penetration level of RES not only results in the degradation of system frequency response [6], but it also increases frequency fluctuation due to variability nature of the RES generation. Thus, frequency regulation becomes much more important for the future low inertia power systems than the traditional power grid with little intermittent renewable generation. Different inverter-based frequency control schematics have been introduced to address frequency stability challenge which can potentially affect the dynamic response of the system in a more active manner. For instance, the synthetic governor control method reserves the wind power generation by making wind turbines work in the over-speed zone instead of maximum power point tracking (MPPT) [7]. Wind power plant inertia control takes advantage of the kinetic energy stored in wind turbines and provides a synthetic inertial frequency response in seconds [8].

These studies [7]-[8] have demonstrated the efficacy of virtual inertia (VI) method which imitates the kinetic inertia of synchronous generator to improve the system dynamic behavior [9]. There are different implementations for synchronous machine response emulation with varying fidelity. Virtual inertia techniques for solar PV generation have been investigated by [10]. It is noted that the virtual inertia requires fast responsive energy buffer; the kinetic energy in a wind turbine and the energy in a battery are limited energy resources for virtual inertia responses.

Traditionally, by looking at the collective performance of all generators using a system equivalent model, a number of performance metrics including frequency nadir and RoCoF are proposed to quantify power system stability. The impact of reduced inertia on system stability has been investigated in [11] where RoCoF and frequency nadir-based constraints are included in the system optimization model. The authors in [12] adopted an approach based on $H_2$ performance metric accounting for the network coherency. However, [13] shows that inertia and frequency response cannot be considered as system-wide magnitudes in power systems. It raised the concern of optimal inertia allocation in the power system.

This paper is to address the aforementioned issues. The contributions of this paper are presented as follows. Explicit models of grid-forming converter (GFC)-based virtual inertia devices are defined to show how such devices contribute to power system frequency stability. In addition, these models are suitable for integration with large scale power system models. In the end, we propose a reduced model based $H_2$-norm algorithm (RMHA) to optimally tune the parameters and the placement of the VI devices in order to enhance the stability of low inertia power systems. The reduced model eliminates passive buses and its Fiedler mode manifests the frequency dynamics of generator buses. Moreover, as a performance metric, the $H_2$-norm based on the proposed reduced model is established beyond the prototypical swing equation.

The remainder of this paper is organized as follows. Section II discusses the power system equivalent model and dynamic model. Section III presents the performance metrics for grid stability and the proposed RMHA method. Section IV shows the simulation results. Section V presents the concluding remarks and future work.

## II. System Frequency Dynamics

### A. System Equivalent Model

The simplified system equivalent model is based on the extension of one-machine swing equation. Applying to an electrical power system that it directly connects rotating machines, the resistance to the change in rotational speed is impacted by the rotating inertia of the rotating mass. The rotating inertia of a synchronous generator is also equal to the stored energy $E_i$ in the rotors of the machine at nominal speed, which is defined as:

$$E_i = \frac{1}{2} J_i {\omega_i}^2 \tag{1}$$

The rotational inertia of a single shaft is commonly defined using its inertia constant given in seconds. It also depends on the rated apparent power [14]. For a single machine, the inertia constant can be expressed as:

$$H_i = \frac{J_i {\omega_i}^2}{2 S_{B_i}} \tag{2}$$

where $H_i$ is the inertia constant of the generator in seconds; $J_i$ is the moment of inertia of the shaft in kg·m$^2$s; $S_{B_i}$ is the base power in MVA; and $\omega_i$ is the nominal rotational speed instead of the actual speed of the machine. It should also be noted that the rotational inertia provided by a single generator is not affected by the actual output power of the generator.

Generators provide rotational inertia to the power system; dynamics of these generators' rotors are directly coupled with the grid electrical dynamics. Thereby the power system could be represented by a single equivalent model of inertia; the total power system inertia $E_{sys}$ is then considered as the summation of the kinetic energy stored in all dispatched generators synchronized with the grid. It can be shown in the form of either the stored kinetic energy or inertia constants as follows.

$$E_{sys} = \sum_{i=1}^{N} \frac{1}{2} J_i {\omega_i}^2 = \sum_{i=1}^{N} H_i S_{B_i} \tag{3}$$

Then the total power rating of the whole power system is represented by

$$S_B = \sum_{i=1}^{N} S_{B_i} \tag{4}$$

The inertia constant of the power system in seconds is given by the equation below,

$$H_{sys} = \frac{\sum_{i=1}^{N} H_i S_{B_i}}{S_B} \tag{5}$$

The swing equation (6) describes the rotor dynamics of the synchronous generator and thereby it also describes the dynamic behavior of the system frequency during a short period of time following a disturbance of power mismatch. For a single generator $i$, the swing equation can be expressed as

$$\frac{\mathrm{d}\omega_i}{\mathrm{d}t} = \frac{P_m - P_{load}}{2 H_i S_{B_i}} \omega_n \tag{6}$$

where $P_m$ is the mechanical power and $P_{load}$ is the load from the power system, while $\omega_n$ is the rated steady state frequency of the system. $\mathrm{d}\omega_i/\mathrm{d}t$ is more commonly known as rate of change of frequency.

The swing equation of the equivalent model can be applied to the whole system. After a disturbance of power mismatch occurrence, the swing equation relates the RoCoF to the total system inertia,

$$\frac{\mathrm{d}\omega}{\mathrm{d}t} = \frac{-\Delta P}{2 H_{sys} S_B} \omega_n \tag{7}$$

where $\Delta P$ is the change in active power in MW.

### B. Dynamic Model

The synchronous generator contributes inertia to the power system through coupled mechanical dynamics of rotor and the electrical dynamics of the whole power system [15]. For a single machine, the dynamic of its rotor can be described in (8) with $M$ and $D$ denoting the normalized inertia and damping constants respectively.

$$P_m - P_e = M \frac{d \triangle \omega}{dt} + D \triangle \omega \tag{8}$$

The transmission network can be considered as a graph consisting of nodes (buses) and edges (branches). Using the topological information and system parameters, the swing equation that describes the single generator dynamics can be applied to all buses to describe the oscillatory behavior of each individual bus [16],

$$m_i \ddot{\theta}_i + d_i \dot{\theta}_i = p_{in,i} - p_{e,i} \tag{9}$$

where $m_i$ and $d_i$ denote the inertia coefficient and damping ratio for node $i$ respectively, while $p_{in,i}$ and $p_{e,i}$ refer to the power input and electrical power output, respectively. Under the assumptions of identical unit voltage magnitudes, the electrical power output at the terminals is related to the voltage phase angles $\{\theta_i\}$ and can be calculated as follows.

$$p_{e,i} = \sum_{j=1}^{n} b_{ij}\left(\theta_i - \theta_j\right), \ \ i \in \{1, \dots, n\} \tag{10}$$

If a bus is a passive load bus, then $m_i$ is considered as zero due to the neglectable contribution to the system frequency dynamics. Since there is no primary droop control on a load bus, the load damping $d_i$ is considered as zero. If a bus is connected to generators, then $m_i$ is the ensemble of nodal generator's rotational inertia and $d_i$ is the nodal droop control coefficient.

## III. METHODOLOGY

In this section, we proposed a novel RMHA method to optimize the allocation of virtual inertia. The effect of Fiedler mode of our proposed reduced model on frequency dynamics are discussed in our studies. Furthermore, the performance metrics considering the Fiedler mode impact are defined to assess the system frequency stability when subjected to a disturbance.

### *A.* Impact of Fiedler Mode

We combine equations (9) and (10) to express the system dynamics, and then the phase angle deviations $\theta$ can be expressed by

$$M\ddot{\theta} + D\,\dot{\theta} = P - L\,\theta \tag{11}$$

where $M$ = diag($\{m_i\}$), $D$ = diag($\{d_i\}$); the nodal power input is represented by vector $P$; and for the Laplacian matrix $L$ of the grid, a linear approximation can be justified considering the angle difference of the voltage phasors are small, giving off-diagonal elements $l_{ij} = -b_{ij}V_i^{(0)}V_j^{(0)}$ and diagonal elements $l_{ii} = \sum_{j=1,j\neq i}^{n} b_{ij}V_i^{(0)}V_j^{(0)}$. The Laplacian matrix is real and symmetric, as such it has a complete orthogonal set of eigenvectors $\{\,u_i\,\}$ with eigenvalues $\{\,\lambda_i\,\}$. The frequency deviations at bus $i$ can be described as follows,

$$\delta\dot{\theta}_i(t) = \frac{\Delta P e^{-\frac{\gamma t}{2}}}{m}\sum_{\alpha=1}^{N} u_{\alpha i}u_{\alpha b}\frac{\sin\left(\sqrt{\frac{\lambda_\alpha}{m}-\frac{\gamma^2}{4}}\,t\right)}{\sqrt{\frac{\lambda_\alpha}{m}-\frac{\gamma^2}{4}}} \tag{12}$$

As $\lambda_1 = 0$ and the first elements of the eigenvectors $u_{\alpha=1,i}$ are all identical, the term associated with $\alpha = 1$ on the right hand side of (12) gives a position-independent contribution to the RoCoF and it is inversely proportional to the inertia coefficient $m$ . The terms associated with $\alpha > 1$ reveal oscillations with both amplitude and period depending on $\sqrt{\lambda_\alpha/m - \gamma^2/4}$ [16]. In this paper, we consider the frequency dynamics of generator buses instead of load buses. By eliminating the passive load buses via Kron reduction [18], we propose a network-reduced power system model with generator buses only and, thus, all the $m$ values are positive for this reduced model. For the reduced model of the IEEE 24-bus system, $\sqrt{\lambda_\alpha/m - \gamma^2/4}$ ranges from [1.29, 20.58] for $\alpha > 1$; high-lying eigenmodes with large $\alpha$ and large eigenvalues $\lambda_\alpha$ contribute much less than low-lying eigenmodes. Hence, the slowest mode among all modes with $\alpha > 1$ is chosen as the metric to measure the regional frequency oscillation amplitude which is called the Fiedler mode [16].

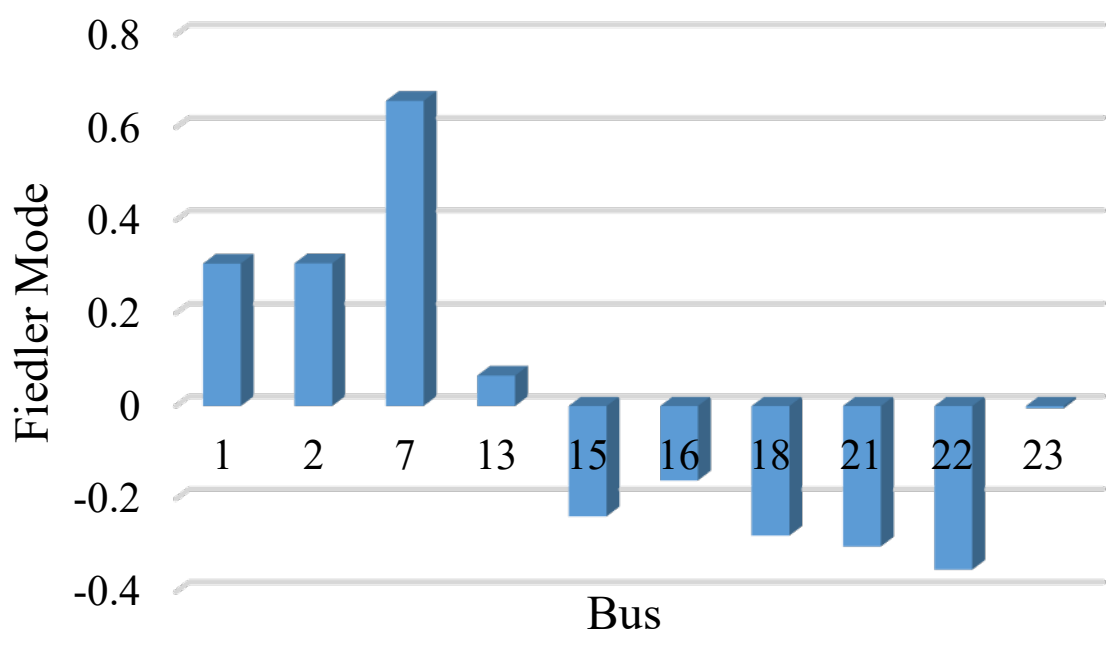

Fig. 1. Fiedler mode distribution.

Fig. 1 shows the distribution of Fiedler mode of the reduced Laplacian matrix of the IEEE 24-bus system. It can be observed that absolute values of Fiedler mode on bus 13 and bus 23 are close to zero, implying relatively less oscillation amplitude for those two buses. Fiedler mode on bus 7 is the largest which indicates larger oscillation. In order to investigate how Fiedler mode affects the frequency response, a contingency case with disturbance on bus 18 is simulated and the frequency responses are shown in Fig. 2. The results show that bus 7 suffers largest frequency oscillation which verifies the previous inference; the frequency on bus 23 is the closest to the center of inertia (COI) frequency response [17], which implies less oscillation.

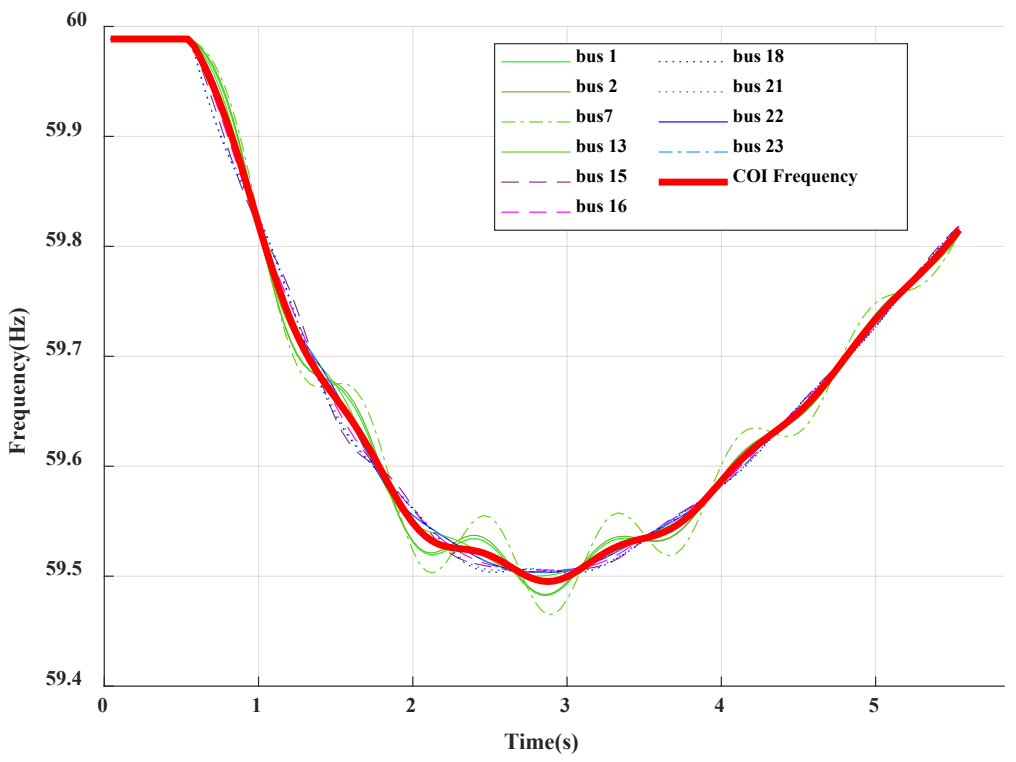

(a) Frequency for period between $t$=0 and $t$=5s.

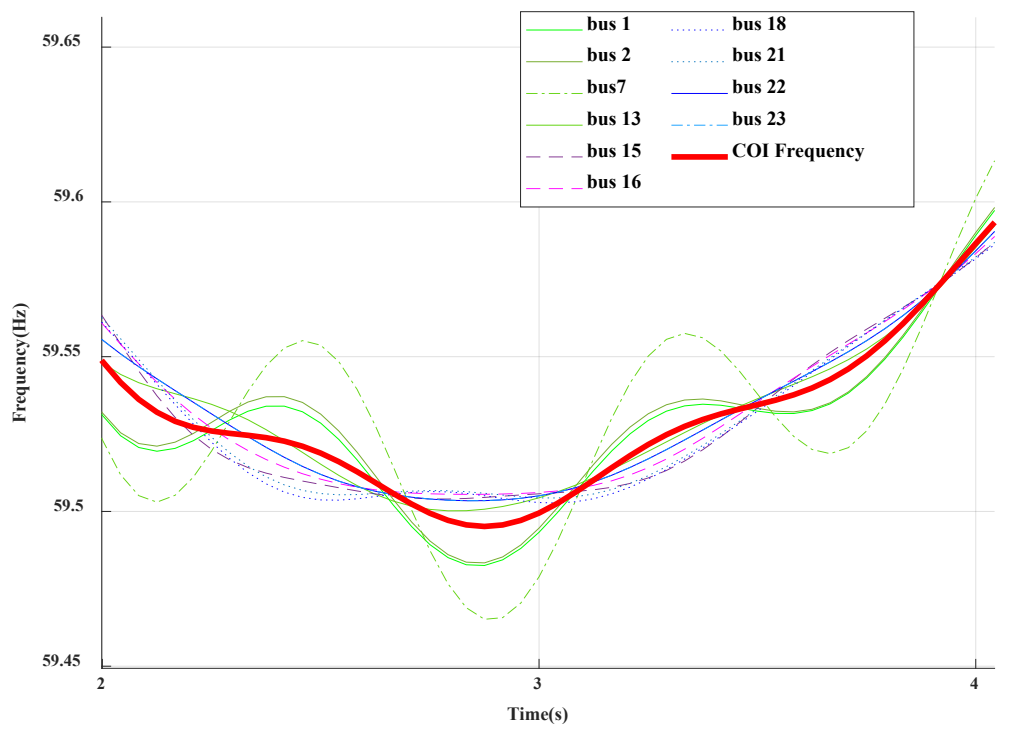

(b) Frequency for the period between $t$=2s and $t$=4s.

Fig. 2. Frequency responses following a disturbance on bus 18.

### *B.* $H_2$-norm Optimization

The implementation of virtual inertia requires active control over GFC-based resources such as distributed RESs and energy storage systems. However, such GFC-based resources are limited and may not be available at each bus to provide virtual inertial responses due to practical conditions. To mitigate frequency stability issues for low-inertia systems, the optimal placement of virtual inertia can be recast as a system norm minimization problem for a linear system with forementioned resources limitations. We consider the linear power system model (11) under a disturbance $p_i$, and the swing equation in state-space vector form is as follows:

$$\begin{bmatrix}\dot{\theta}\\ \dot{\omega}\end{bmatrix} = \begin{bmatrix}0 & I\\ -M^{-1}L & -M^{-1}D\end{bmatrix}\begin{bmatrix}\theta\\ \omega\end{bmatrix} + \begin{bmatrix}0\\ M^{-1}\end{bmatrix}p_i \tag{13}$$

where $\omega = \dot{\theta}$ and $\dot{\omega} = \ddot{\theta}$. The system (13) can also be written in standard state-space form:

$$\dot{x} = A \cdot x + B \cdot u \tag{14}$$

The grid-forming VI devices are power electronic devices that mimic the inertial response of synchronous generators. Grid-forming converter uses a voltage source connected to the grid via an LC filter with parasitic losses. Hence the virtual inertia device is modeled as:

$$\widetilde{m}_{VI}\dot{\omega}_{VI} = -\tilde{d}_{VI}\omega_{VI} - P_{VI} \tag{15}$$

where $P_{VI}$ is the active power from the grid-forming VI device into the grid; $\widetilde{m}_{VI}$ is the configurable virtual inertia coefficient; and $\tilde{d}_{VI}$ is the virtual damping constant. These two variables are constrained by realistic values which ensure the system frequency in the normal operating regime. Augmenting the VI control to the system, the updated system representation is as follows,

$$\begin{bmatrix}\dot{\theta}\\ \dot{\omega}\end{bmatrix} = \begin{bmatrix}0 & I\\ -{M_{cl}}^{-1}L & -{M_{cl}}^{-1}D_{cl}\end{bmatrix}\begin{bmatrix}\theta\\ \omega\end{bmatrix} + \begin{bmatrix}0\\ {M_{cl}}^{-1}\end{bmatrix}p_i \tag{16}$$

where $M_{cl} = M + M_{VI}$ and $D_{cl} = D + D_{VI}$. In order to model the disturbance in the power system, the input $p_i$ is remodeled as $p_i = V^{\frac{1}{2}}v$, where $v$ denotes the disturbance input and $V$ is a diagonal matrix describing the disturbance magnitude and locations. Therefore, the final state space model is expressed as,

$$\begin{bmatrix}\dot{\theta}\\ \dot{\omega}\end{bmatrix} = \underbrace{\begin{bmatrix}0 & I\\ -{M_{cl}}^{-1}L & -{M_{cl}}^{-1}D_{cl}\end{bmatrix}}_{=A}\begin{bmatrix}\theta\\ \omega\end{bmatrix} + \underbrace{\begin{bmatrix}0\\ {M_{cl}}^{-1}V^{\frac{1}{2}}\end{bmatrix}}_{=B}v \tag{17}$$

Based on the model presented above, we propose a novel performance metric to assess the frequency stability of the grid: Fiedler mode weighted coherency index (FMWCI) that penalizes angular differences and frequency excursions. The proposed performance metric FMWCI is defined in (18), which includes a quadratic term of angle differences and a quadratic term of frequency displacements.

$$\int_0^{\infty}\left\{\sum_{i,j=1}^{n} b_{ij}\left(\theta_i(t) - \theta_j(t)\right)^2 + \sum_{i=1}^{n} u_i\omega_i^2(t)\right\}dt \tag{18}$$

Here, $u_i$ is the absolute value of the Fiedler mode of the network Laplacian matrix. The magnitude of RoCoF and frequency deviations strongly depend on the Fiedler mode absolute value; hence the positive scalars $u_i$ are considered as error penalization of frequency displacements.

Adopting the state representation (17), the performance metric FMWCI (18) would be equal to the time-integral of the performance matrix $C$ that is defined as follow:

$$y = \underbrace{\begin{bmatrix}N_R^{\frac{1}{2}} & 0\\ 0 & U^{\frac{1}{2}}\end{bmatrix}}_{=C}\begin{bmatrix}\theta\\ \omega\end{bmatrix} \tag{19}$$

The observability Gramian of the system is defined as,

$$P = \int_0^{\infty} e^{A^T t} \cdot C^T \cdot C \cdot e^{At} dt \tag{20}$$

$P$ is uniquely defined as a solution of the Lyapunov equation (21).

$$PA + A^T P + C^T C = 0 \tag{21}$$

For the state-space system defined above, by solving (21) for $P$, we have the $H_2$-norm computed as follows:

$$\|\mathcal{g}\|_2 = \sqrt{Trace\ (B^T PB)} \tag{22}$$

For a given virtual inertia budget $M_{budget}$, the objective is to find the configuration of VI device control coefficient, which minimizes the $H_2$-norm. This problem is summarized as:

$$\underset{m_i}{\text{minimize}} \quad \|\mathcal{g}\|_2^2 = Trace\ (B^T PB) \tag{23}$$

subject to:

$$PA + A^T P + C^T C = 0 \tag{23a}$$

$$m_i^{VI} + \underline{m_i} = m_i \in \left[\underline{m_i}, \overline{m_i}\right],\ i \in \{1, \ldots, n\}, \tag{23b}$$

$$\sum_{i\in N} m_i^{VI} = M_{budget}, \tag{23c}$$

where $\underline{m_i}$ is the lower bound of inertia constant at bus $i$ accounting for the inertia of dispatched synchronous generators; $\overline{m_i}$ is the upper bound of inertia constant at bus $i$ accounting for both the available virtual inertia and synchronous inertia; $m_i^{VI}$ denotes the virtual inertia part of distributed nodal inertia. Gradient-based methods are used to directly optimize the inertia constants of a linearized networked swing equation model to minimize the $H_2$-norm of the power system.

## IV. CASE STUDIES

The numerical simulations were conducted on the IEEE 24-bus system [19]. The base system contains 24 buses, 33 generators and 38 lines. The total generation capacity from synchronous generators is 4,606 MW and the system peak load is 3,461 MW.

### *A. Validation of VI Devices*

To investigate the impact of VI devices on power system frequency stability, two cases are conducted in MATLAB Simulink. Initially, the power system is operating in steady state. To validate the linearized model, we compare it to the non-linear model for step disturbances. The step disturbance is applied on 23 respectively which is set as -150 MW, or a load increase of 150 MW.Fig. 3 and Fig. 4 show the frequency evolution of synchronous generators in the case without VI devices and in the case with VI devices respectively. It is observed that under the disturbance on bus 23, without implementation of VI devices, the frequency nadir of synchronous generator reaches below 59.84 Hz and the highest RoCoF is 0.7 Hz/s. For the same disturbance on bus 23, the frequency nadir and RoCoF are significantly improved with VI devices by comparing Fig. 3 and Fig. 4. The highest frequency deviation is 0.22 Hz in the case without VI devices while it is only 0.21 Hz when a VI device providing only 100 MWs non-synchronous inertia is included in the system.

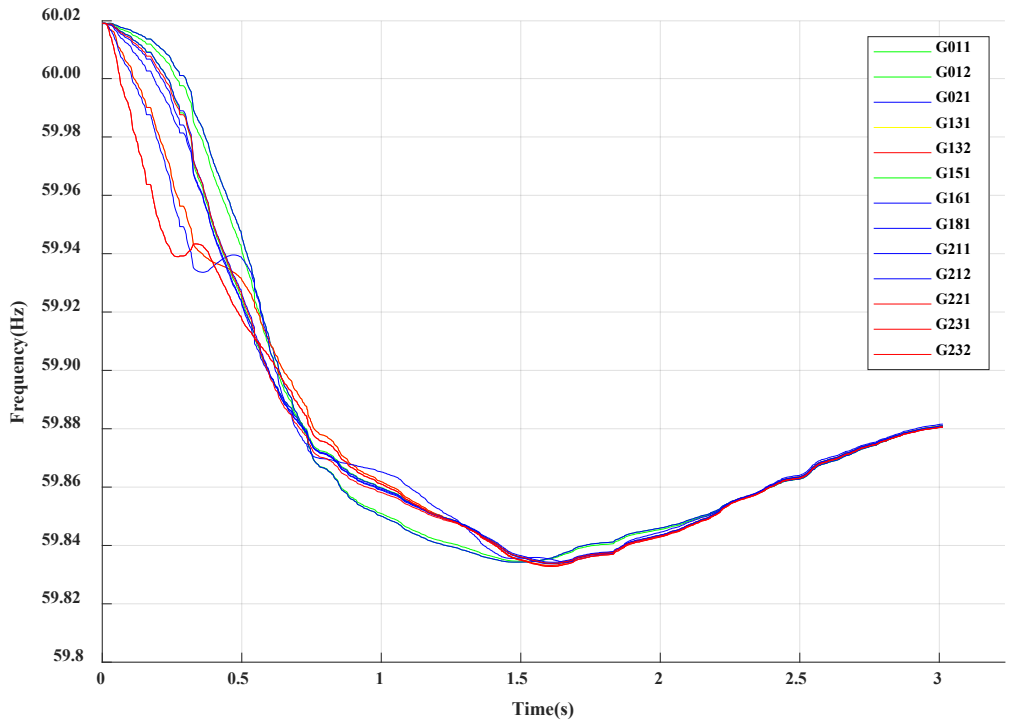

Fig. 3. Frequency evolution of synchronous generators following a disturbance without GFC-based VI devices.

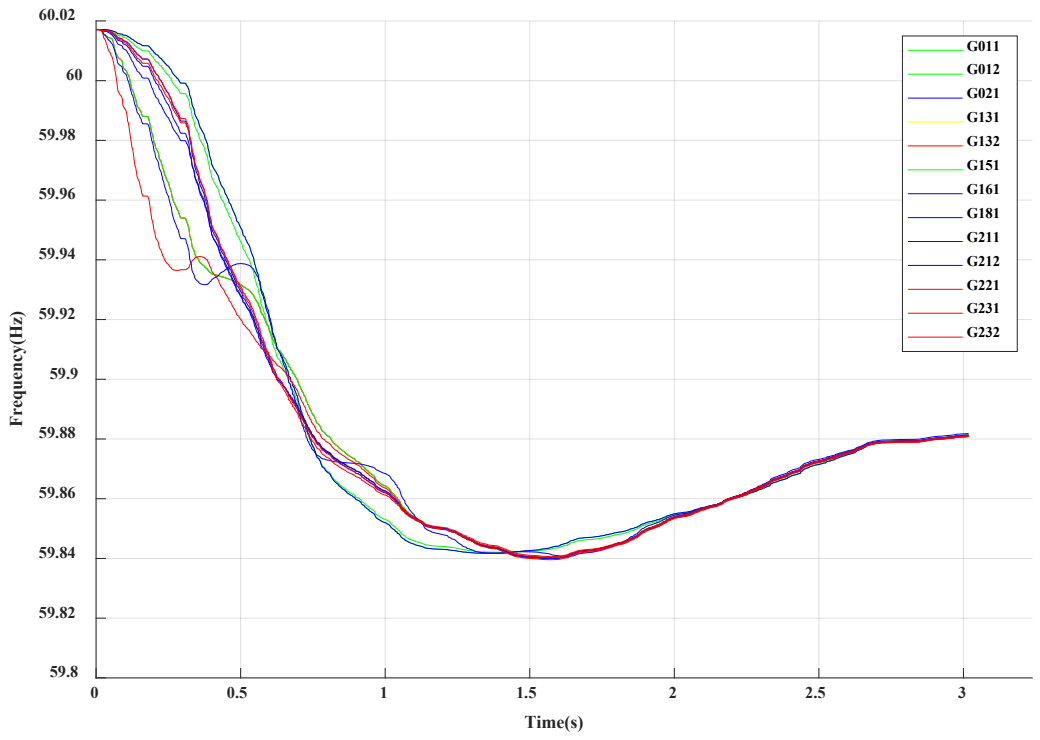

Fig. 4. Frequency evolution of synchronous generators following a disturbance with GFC-based VI devices.

Fig. 5 presents the distribution of post-disturbance frequency nadirs of all generators for the case without VI devices and the case with VI devices installed on bus 23. It can be concluded that the mean and variance of the distribution of frequency nadirs is smaller for the system equipped with the VI device as compared to the system without VI device. Therefore, we can conclude that VI devices have the expected positive impact on frequency stability.

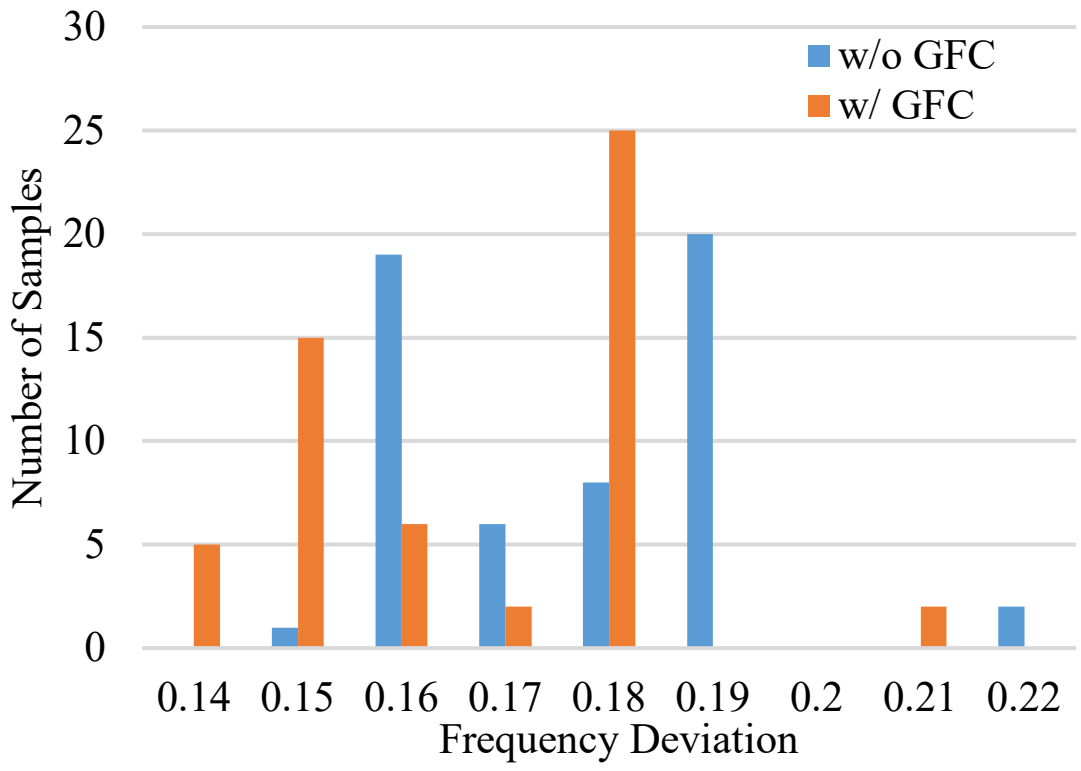

Fig. 5. Distributions of generator frequency deviation with and without GFC.

## B. *Optimal Placement of VI Devices*

The optimal inertia profile for the system is obtained by solving the optimization problem (23). The resulting optimal inertia allocation is depicted in Fig. 6. It can be observed that the largest allocation of virtual inertia is on bus 7 and virtual inertia allocated on bus 23 is much less. This indicates that inertia is mostly needed on buses with higher Fiedler mode values to enhance system frequency stability, which is in agreement with the findings in Section III.

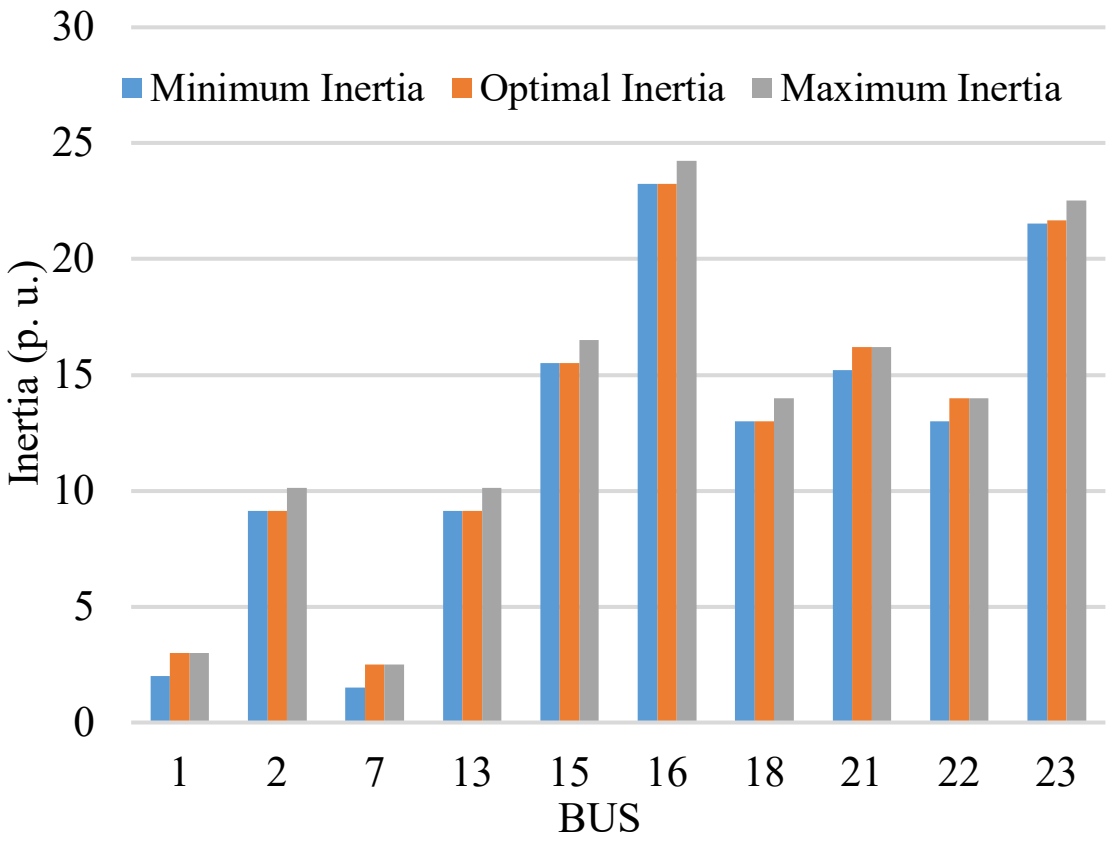

Fig. 6. Optimal inertia allocation for IEEE 24-bus system.

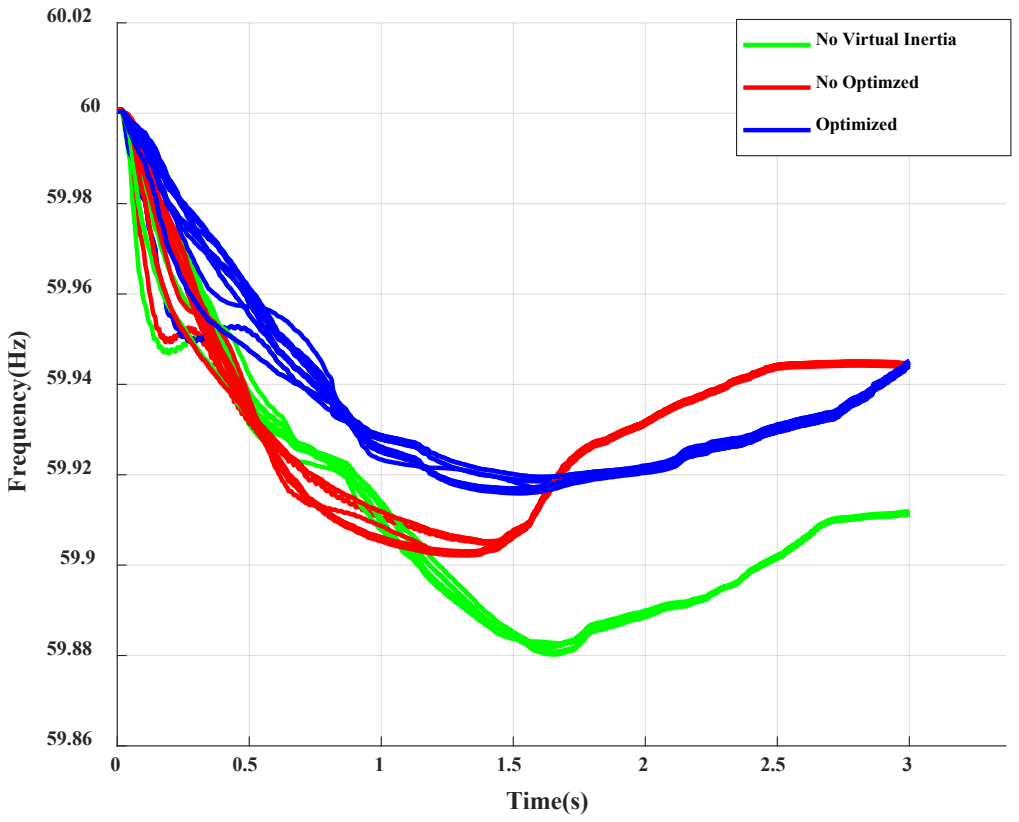

Fig. 7. Frequency responses with and without virtual inertia.

Three cases with different virtual inertia configurations are performed following a total step increase of 150MW in load. The case with no VI implemented is set as the base case; virtual inertia devices are utilized in case 2 while distribution is not optimized; same total value of virtual inertia is implemented in case 3, and the optimal allocation of virtual inertia is achieved by configuring control coefficient of VI devices based on the proposed RMHA. As shown in Fig. 7 and Table I, the results of no virtual inertia, no optimized virtual inertia and optimal virtual inertia allocation are compared. The case with optimal virtual inertia allocation reduces the maximal RoCoF from -0.22 Hz/s to -0.11 Hz/s comparing to the base case, while the system suffers higher RoCoF in the case with no optimized virtual inertia allocation. The results also show that the optimized virtual inertia improves the average frequency nadir from 59.88 Hz to 59.92 Hz, while the arresting time is shortened by 0.69s (32%) from 2.18s to 1.49s. This is a significant improvement, which allows more time for the system to respond.

TABLE I Simulation results with and without virtual inertia

| Case | Average RoCoF [Hz/s] | Frequency nadir [Hz] | Time to nadir [s] |
|---|---|---|---|
| No VI | -0.22 | 59.87 | 2.02 |
| No optimized VI | -0.17 | 59.90 | 1.26 |
| Optimized VI | -0.11 | 59.92 | 1.37 |

The power injections from the VI devices are plotted in Fig. 8. The grid-forming VI devices respond to the disturbance event instantaneously, which immediately helps mitigating the initial RoCoF and reducing the frequency excursion on the synchronous generators. Fig. 8(b) shows the power injection of devices with same control coefficient, it can be observed that even with the same configured parameters, the power outputs of the VI devices vary substantially due to their relative locations to the disturbance. This indicates that the devices in different regions take distinct actions under the same disturbance.

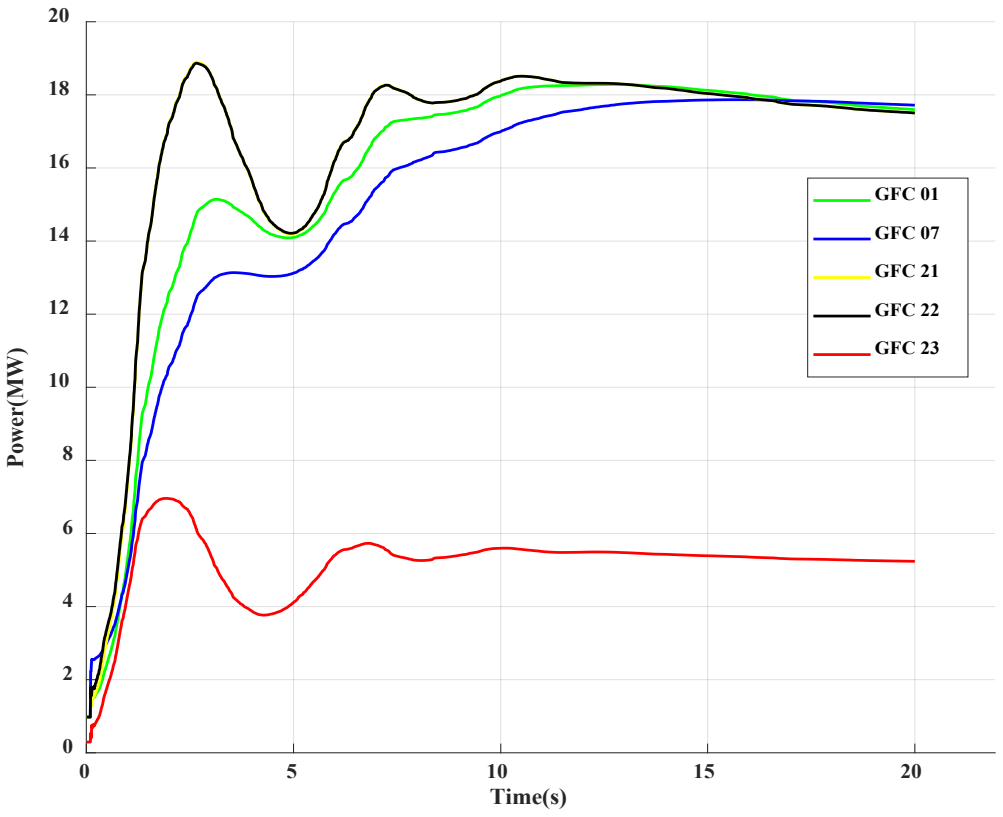

(a) Time period from $t$=0 to $t$=20s.

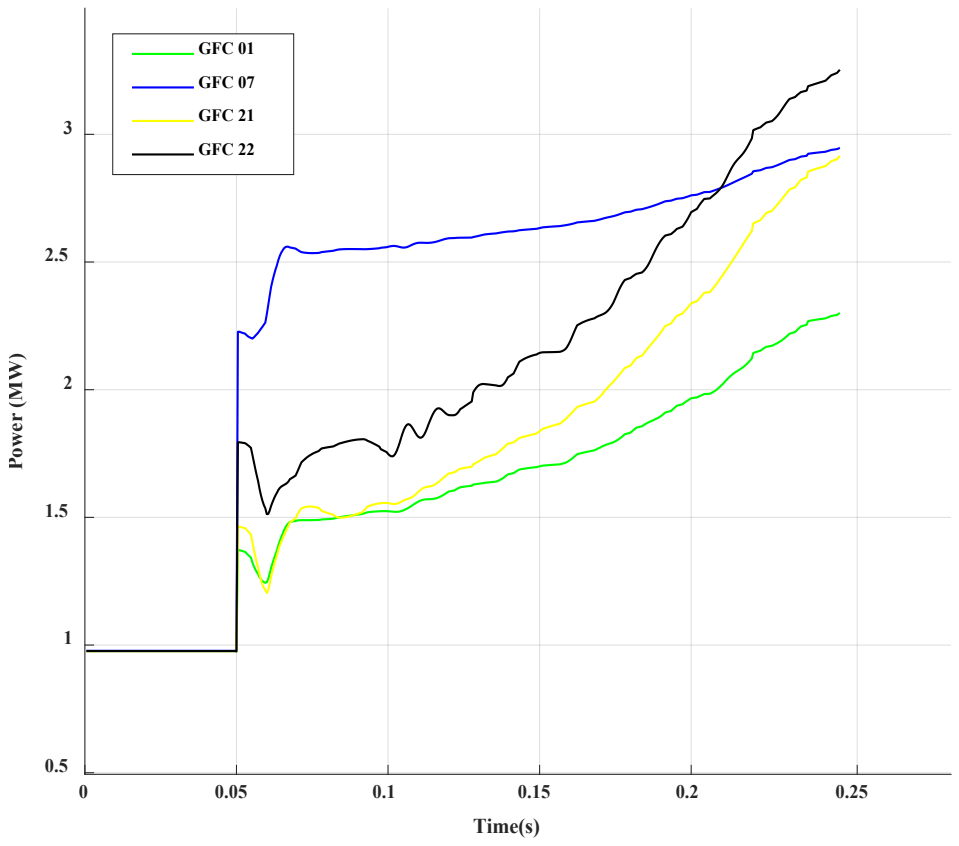

(b) Time period from $t$=0 to $t$=0.25s.

Fig. 8. Power injections of VI devices with same inertia contribution.

## V. Conclusions

In this paper, we comprehensively examine the dynamics of low-inertia power systems with virtual inertia devices. A brief investigation into regional frequency responses corresponding to the Fiedler modes points out various options to improve system inertia response, including virtual inertia control and optimal inertia allocation. We implement grid-forming virtual inertia devices as frequency dynamic controller to provide extra inertia. The RMHA method proposed in this paper optimally tune the parameters and the placement of the VI devices by incorporating the impact of Fiedler mode.

The simulation results on the IEEE 24-bus system indicate that the implementation of VI devices improve the stability of power system frequency responses in terms of frequency nadir and maximum RoCoF. Our proposed RMHA method has been proved to enhance the frequency dynamics of generator buses. Further lines of work on this topic should explore how to include probabilistic metrics for the provision of frequency response. Allocation and acquirement of virtual inertia can consider more economic factors.